\documentclass[final,5p,times,twocolumn]{elsarticle}

\usepackage{graphicx}
\usepackage{epsfig}
\usepackage{amssymb}
\usepackage{amsmath}
\usepackage{slashed}
\usepackage{units}

\biboptions{sort&compress}

\journal{Physics Letters B}

\begin{document}

\begin{frontmatter}

\title{Experimental limits on neutron disappearance into another braneworld}

\author[ms]{Micha\"{e}l Sarrazin\corref{cor}}
\ead{michael.sarrazin@fundp.ac.be}
\author[gp]{Guillaume Pignol}
\ead{guillaume.pignol@lpsc.in2p3.fr}
\author[fp]{Fabrice Petit}
\ead{f.petit@bcrc.be}
\author[vvn]{Valery V. Nesvizhevsky}
\ead{nesvizhevsky@ill.eu}

\cortext[cor]{Corresponding author}

\address[ms]{Department of Physics, University of Namur (FUNDP), 61 rue de Bruxelles, B-5000 Namur, Belgium}
\address[gp]{LPSC, Universit\'{e} Joseph Fourier, CNRS/IN2P3 - INPG, 53 rue des Martyrs, F-38026 Grenoble,
France}
\address[fp]{BCRC (Member of EMRA), 4 avenue du gouverneur Cornez, B-7000 Mons, Belgium}
\address[vvn]{Institut Laue-Langevin, 6 rue Jules Horowitz, F-38042 Grenoble, France}

\begin{abstract}
Recent theoretical works have shown that matter swapping between two
parallel braneworlds could occur under the influence of magnetic vector
potentials. In our visible world, galactic magnetism possibly produces a
huge magnetic potential. As a consequence, this paper discusses the
possibility to observe neutron disappearance into another braneworld in
certain circumstances. The setup under consideration involves stored
ultracold neutrons $-$ in a vessel $-$ which should exhibit a non-zero
probability $p$ to disappear into an invisible brane at each wall collision.
An upper limit of $p$ is assessed based on available experimental results.
This value is then used to constrain the parameters of the theoretical
model. Possible improvements of the experiments are discussed, including enhanced stimulated swapping by artificial means.
\end{abstract}


\end{frontmatter}

\section{Introduction}

According to the braneworld hypothesis, our observable Universe can be
considered as a three-dimensional space sheet (a $3$-brane) embedded in a
larger spacetime with $N>4$ dimensions (the bulk) \cite{Review}. Brane is a
concept inherited from high energy physics and unification models. Testing
the existence of branes or extra dimensions is therefore becoming a
fundamental challenge. Such evidences are expected to be obtained through
high energy collisions \cite{Pheno,bbb,HEP}, but it has been also
demonstrated that some detectable effects could also be observed at low
energy \cite{Pheno,bbb,Gravi,Mirrornu,Tunnel,Mirrorphot1,Mirrorphot2,Mirrorphot3,Mirrorneut,MirrorPosi,MirrorPosi2,hidden,BraneNC,PLB,Reso,fcomb,IJMPA,neule}
. This is the topic of the present paper.

Some authors have early underlined or suggested that the particles of the
standard model could be able to escape out of our visible world \cite
{Pheno,bbb,Mirrornu}. Many effects can be considered and have been explored
until now along this line of thought. For instance, in some approaches,
particles are expected to leak into the bulk through a tunnel effect \cite
{Tunnel}. Other works also considered that fluctuations of our home-brane
could yield small bubbles branes, which carry chargeless matter particles
(such as neutrons for instance) into the bulk \cite{bbb}. In another context,
other approaches consider some coupling between certain particles of the
standard model and some hidden or dark sectors \cite
{Mirrornu,Mirrorphot1,Mirrorphot2,Mirrorphot3,Mirrorneut,MirrorPosi,MirrorPosi2,hidden}.
It is sometimes suspected that such hidden sectors could live in other
branes. It is the case with the photon-hidden photon kinetic mixing \cite{Mirrorphot1,Mirrorphot2,Mirrorphot3}. A $U(1)$\ field on a hidden brane can
be coupled to the $U(1)$\ photon field of our brane through a one-loop
process in a stringy context \cite{Mirrorphot2,Mirrorphot3}. In the mirror
world approaches, the matter-mirror matter mixing is
also considered (with neutron and mirror neutron \cite{Mirrorneut} for instance) though, in the best of our knowledge, a full
derivation through a brane formalism is still lacking. Actually, ultracold
neutron (UCN) experiments related to the neutron disappearance are then
fundamental since they could allow to quantify or to distinguish among the
different predicted phenomenologies \cite{neuexp1,neuexp2}.

In previous works \cite{BraneNC,PLB,Reso,fcomb,IJMPA}, two of the present
authors (Sarrazin and Petit) have shown that for a bulk containing at least
two parallel 3-branes hidden to each other, matter swapping between these
two worlds should occur. The particle must be massive, can be electrically charged\footnote{The model used in the present work can be proved and
derived \cite{BraneNC} from a domain wall approach in which the Dvali-Gabadadze-Shifman
mechanism \cite{DGS} is responsible for the gauge field localization on the
brane. This mechanism allows electric charge leaks \cite{DGS} contrary to
other models, such as the bubble brane approach \cite{bbb} for instance.} or not, but must be endowed with a magnetic moment. This
swapping effect between two neighboring 3-branes is triggered by using
suitable magnetic vector potentials. More important, this new effect $-$
different from those previously described in literature $-$ could be
detected and controlled with present day technology which opens the door to
a possible experimental confirmation of the braneworld hypothesis. For
charged particles, the swapping is possible though a few more difficult to
achieve \cite{Reso}. As a consequence, for a sake of simplicity and in order
to be able to distinguish the swapping effect with other kind of predicted
phenomena, we suggested the use of neutron for a prospective experiment.

In the present work we discuss the possibility that an astrophysical
magnetic vector potential could lead to such a matter swapping. The basic
argument is that the astrophysical vector potentials are considerably larger
than any other counterpart generated in a laboratory. A
possible consequence for free neutrons would be then high frequency and
small amplitude oscillations of the matter swapping probability between the
two branes. Ultracold neutrons stored in a vessel would therefore have a non-zero probability $p$ to escape from our brane toward the hidden brane at
each wall collision. Such a process would be perceived as a neutron
disappearance from the point of view of an observer located in our brane.
The purpose of this paper is to assess an upper limit on $p$ based on
already published data in literature. This upper limit is then used to
constrain the parameters of the model. On the basis of this assessment, more
sensitive experiments are suggested and described.

In section \ref{model}, the model describing the low-energy dynamics of a neutron in
a two-brane Universe is recalled. The conditions leading to matter swapping
between branes are given. We discuss the origin $-$ and the magnitude $-$ of
the ambient magnetic vector potential, which is required to observe matter
exchange between branes. The gravitational environment that can impede the
swapping to occur, is also discussed. In section \ref{sec3}, available data from
literature are analyzed and used to constrain the parameters of the
two-brane Universe model. Finally, in section \ref{further} improvements of the
experimental setup are suggested. A variable-collision-rate experiment is
proposed. A long timescale experiment as well as a laser-induced matter
swapping experiment are also discussed.

\section{Matter swapping between two braneworlds}

\label{model}

In previous works \cite{BraneNC,PLB}, it was shown that in a Universe
containing two parallel braneworlds invisible to each other, the quantum
dynamics of a spin$-1/2$ fermion can be described by a two-brane Pauli
equation at low energies. For a neutron outside a nucleus, in
electromagnetic and gravitational fields, the relevant equations can be
written as \cite{BraneNC}:
\begin{equation}
i\hbar \frac \partial {\partial t}\left(
\begin{array}{c}
\psi _{+} \\
\psi _{-}
\end{array}
\right) =\left\{ \mathbf{H}_0+\mathbf{H}_{cm}\right\} \left(
\begin{array}{c}
\psi _{+} \\
\psi _{-}
\end{array}
\right)   \label{2BPauli}
\end{equation}
where the indices $\pm $ are purely conventional and simply allow to
discriminate the two branes. $\psi _{+}$ and $\psi _{-}$ are usual Pauli
spinors corresponding to the wave functions in the $(+)$ and $(-)$ branes
respectively, and where
\begin{equation}
\mathbf{H}_0=\left(
\begin{array}{cc}
\mu \mathbf{\sigma \cdot B}_{+}+V_{+} & 0 \\
0 & \mu \mathbf{\sigma \cdot B}_{-}+V_{-}
\end{array}
\right)   \label{UsualPauli}
\end{equation}
and
\begin{eqnarray}
\mathbf{H}_{cm}=-ig\mu \left(
\begin{array}{cc}
0 & \mathbf{\sigma \cdot }\left\{ \mathbf{A}_{+}-\mathbf{A}_{-}\right\}  \\
-\mathbf{\sigma \cdot }\left\{ \mathbf{A}_{+}-\mathbf{A}_{-}\right\}  & 0
\end{array}
\right)   \label{Coupling}
\end{eqnarray}
such that $\mathbf{A}_{+}$ and $\mathbf{A}_{-}$ correspond to the magnetic
vector potentials in the branes $(+)$ and $(-)$ respectively. The same
convention is applied to the magnetic fields $\mathbf{B}_{\pm }$ and to the
gravitational potentials $V_{\pm }$. $\mu $ is the magnetic moment of the
particle. Each diagonal term of $\mathbf{H}_0$ is simply the usual Pauli
Hamiltonian for each brane. In addition to these usual terms, the two-brane
Hamiltonian comprises also a new term $\mathbf{H}_{cm}$ fully specific of a
Universe with two branes \cite{BraneNC}. $\mathbf{H}_{cm}$ implies that
matter exchange between branes depends on the magnetic moment and on the
difference between the local (i.e. on a brane) values of the magnetic vector
potentials. $\mathbf{H}_{cm}$ leads to a phenomenology which shares some
similarities with the mirror matter paradigm \cite{Mirrornu,Mirrorneut,MirrorPosi,MirrorPosi2} or
other approaches involving some hidden or dark matter sectors \cite{hidden}. Nevertheless the present approach differs in several ways:

\begin{itemize}
\item  In the mirror (or hidden) matter formalism, it is often
considered by implication that only one four-dimensional manifold exists and
that particles split into two families: the standard and the mirror sectors.
By contrast, in the present model, particles are restricted to those of the
standard model but they have access to two distinct $3$-branes (i.e. two
distinct four-dimensional manifolds). As a consequence, in the mirror matter
approach, matter and mirror matter particles undergo the same gravitational fields
(i.e. $V_{+}=V_{-}$)
\footnote{Note that, in a recent paper \cite{bigrav} and in a
different context related to the dark matter gravity, Berezhiani \textit{et
al} suggested that matter and mirror matter could not necessarily feel
exactly the same gravitational interaction in the context of a bigravity
approach.} whereas in the present braneworld
approach, matter is subjected to the gravitational fields of each brane
(i.e. $V_{+}\neq V_{-}$) that possess their own gravitational sources.

\item  The coupling between the particles of each brane occurs in a specific
way without recourse of the coupling considered in the mirror matter
concept. The coupling term $\mathbf{H}_{cm}$ is specific to the braneworld
approach \cite{BraneNC}. Since its value can be changed by modifying the
local value of the magnetic vector potential, it can be consequently
controlled by artificial means. The effects related to $\mathbf{H}_{cm}$ \cite{fcomb,IJMPA} can
be then distinguished from those related to coupling from other models \cite
{Mirrorneut,MirrorPosi}.
\end{itemize}

However, instead of considering artificial (i.e. generated in a laboratory)
vector potentials we are now focusing on the case of an ambient magnetic
potential with an astrophysical origin (section \ref{Magvect}). Let $\mathbf{%
A}_{amb}=\mathbf{A}_{amb,+}-\mathbf{A}_{amb,-}$ be the difference of ambient
magnetic potentials of each brane\footnote{Usually, assessment of the value and direction of this
cosmological magnetic potential field is ambiguous because it has no
physical meaning in non-exotic gauge invariant physics. But in the present context
the difference of the field between both branes would become physical.}. Assuming that $\mu B_{\pm }\ll V_{\pm }$,
i.e. one can neglect the magnetic fields in the branes (especially one
assumes that $\mathbf{\nabla }\times \mathbf{A}_{amb}=\mathbf{0}$), then by
solving the Pauli equation, it can be shown that the probability for a
neutron initially localized in our brane to be found in the other brane is:
\begin{equation}
P=\frac{4\Omega ^2}{\eta ^2+4\Omega ^2}\sin ^2\left( (1/2)\sqrt{\eta
^2+4\Omega ^2}t\right)   \label{Rabi}
\end{equation}
where $\eta =|V_{+}-V_{-}|/\hbar $ and $\Omega =g\mu A_{amb}/\hbar $. Eq. (\ref{Rabi}) shows that the neutron in the potential $A_{amb}$ undergoes
Rabi-like oscillations between the branes. Note that the probability does
not depend on the neutron spin state by contrast with the magnetic vector
potential direction in space. As detailed in previous papers \cite
{fcomb,Reso}, the environmental interactions (related to $V_{\pm }$) are
usually strong enough to almost suppress these oscillations. This can be
verified by considering Eq. (\ref{Rabi}) showing that $P$ decreases as $\eta
$ increases relatively to $\Omega $. Since no such oscillations have been
observed so far, we can assume that the ratio $\Omega /\eta $ is usually
very small. As a consequence, the oscillations present a weak amplitude and
a high angular frequency of the order $\eta /2$. Let us now consider a
neutron gas in a vessel. In the general case, it is in a superposition of
two states: neutron in our brane vs. neutron in the other brane. A collision
between the neutron and a wall of the vessel acts therefore as a measurement
and the neutron collapses either in our brane with a probability $1-p$ or in
the other invisible brane with a probability $p$. It is therefore natural to
consider that the neutron swapping rate $\Gamma $ into the other brane is
related to the collision rate $\gamma $ between the neutron and the vessel
walls, i.e.
\begin{equation}
\Gamma =\gamma p  \label{6}
\end{equation}
with $\gamma =1/\left\langle t_f\right\rangle $ where $\left\langle
t_f\right\rangle $ is the average flight time of neutrons between two
collisions on vessel walls. The probability $p$ is the average of $P$
according to the statistical distribution of the free flight times, i.e.
\begin{equation}
p=\left\langle P\right\rangle \sim \frac{2\Omega ^2}{\eta ^2}  \label{Proba}
\end{equation}

Equations (\ref{6}) and (\ref{Proba}) are valid provided a large number of
oscillations occur during a given time interval. This must be verified for $%
\left\langle t_f\right\rangle $ (i.e. $\eta \gg \gamma $) but also during
the duration $\delta t$ of a neutron collision on a wall (i.e. $\eta \gg
1/\delta t$) \footnote{$\delta t$ can be estimated as the time needed for a
neutron to make a round trip along the penetration depth $d$ of the wall
(typically $d\approx 10$ nm \cite{Ignatovich}). With $\delta t\sim 2d/v$ (where $\nu $ is the
UCN velocity, here $\nu \approx 4$ m$\cdot $s$^{-1}$), one gets $%
\delta t\approx 5$ ns.}. As shown hereafter, the lowest considered energies $%
\eta \hbar $ are about $10^{-2}$ eV, i.e. $\eta \approx 2\times 10^{13}$ Hz,
while $1/\delta t\approx 2\times 10^8$ Hz and the greatest values of $\gamma $
are about $20$ Hz. As a consequence, the Eqs. (\ref{6}) and (\ref{Proba})
are legitimate in the present context.

\subsection{Ambient magnetic vector potential}

\label{Magvect}

Let us now consider that a natural astrophysical magnetic vector potential $%
\mathbf{A}_{amb}$ may have on the neutron dynamics in the vessel. The magnitude of such a potential was recently discussed in
literature since it allows to constrain the photon mass \cite
{vectpot,vectpot2}. Of course, in the present work, we are not concerned by
such exotic property of photon, which is assumed to be massless. $\mathbf{A}%
_{amb}$ corresponds typically to a sum of individual contributions coming
from astrophysical objects (star, planet, galaxy,...) that surround us.
Indeed, each astrophysical object endowed with a magnetic moment $\mathbf{m}$
induces a magnetic potential:
\begin{equation}
\mathbf{A}(\mathbf{r})=\frac{\mu _0}{4\pi }\frac{\mathbf{m}\times \mathbf{r}%
}{r^3}  \label{MagPot}
\end{equation}
from which the magnetic field $\mathbf{B}(\mathbf{r})=\mathbf{\nabla }\times
\mathbf{A}(\mathbf{r})$ produced by the object can be deduced. We get $A\sim
RB$ where $R$ is the typical distance from the astrophysical source. In the
vicinity of Earth for instance, and at large distances from sources, we note
that $\mathbf{A}$ is then almost constant (i.e. $\mathbf{\nabla }\times
\mathbf{A}_{amb}\approx \mathbf{0}$) and cannot be canceled with magnetic
shields \cite{vectpot}. The magnitude of the main expected contributions to $%
\mathbf{A}_{amb}$ can be easily deduced. For instance, the Earth
contribution to $A_{amb}$ is about $200$ T$\cdot $m while it is only $10$ T$%
\cdot $m for the Sun \cite{vectpot}. Now, if we consider the galactic
magnetic field \textbf{(}about $1$\textbf{\ }$\mu $G \cite{GalMag}\textbf{) }%
relatively to the Milky Way core \textbf{(}at\textbf{\ }about $1.9\times
10^{19}$\textbf{\ }m\textbf{)} one gets $A\approx 2\times 10^9$\ T$\cdot $m
\cite{vectpot,vectpot2}. Note that in Refs. \cite{vectpot} a value of $A\approx
10^{12}$ T$\cdot $m was derived for the Coma galactic cluster.
Unfortunately, other authors (see Goldhaber and Nieto in Ref. \cite{vectpot2}
for instance), underlined that substantial inhomogeneities can exist in the
field distributions such that $A_{amb}$ may strongly vary in different
regions. As a consequence, there is a lack of knowledge concerning the
magnitude of contributions from extragalactic scales \cite{vectpot2, GalMag2}%
. Therefore the value given in \cite{vectpot} cannot be presently considered
as reliable enough to be used here. In addition, Eq. (\ref{Coupling}) shows
that it is the difference $\mathbf{A}_{amb}=\mathbf{A}_{amb,+}-\mathbf{A}%
_{amb,-}$ between the vector potentials of the two braneworlds that is
relevant. Since $\mathbf{A}_{amb,-}$ depends on unknown sources in the other
brane, we cannot assess its value. For all these reasons, we should consider
$\mathbf{A}_{amb}$ as an unknown parameter of the model. Nevertheless, we
will admit that a value of $A_{amb}\approx 2\times 10^9$\ T$\cdot $m is
probably of the right order of magnitude \cite{vectpot,vectpot2}.

\subsection{Gravitational potential $\eta$}

\label{envpot}

In the present context, $\eta =|V_{+}-V_{-}|/\hbar$ and only gravitational
interactions are relevant. It is difficult to specify the value of $\eta
\hbar $. Indeed, since the gravitational contribution of the hidden world ($%
V_{-} $) is unknown, $\eta$ must be therefore an unknown parameter of the
model. However, according to the estimations given in previous works \cite
{Reso}, $V_{+}$ could be of the order of $500$ eV due to the Milky Way core
gravitational potential acting on neutron. By contrast, the Sun, the Earth
and the Moon contribute for about $9$ eV, $0.65$ eV and $0.1$ meV. As a
consequence, one can fairly suppose that the value of $\eta$ is included in
a range from few meV up to few keV.

At last, one notes that $\eta$ must be also time-dependent. Let us consider
the significant motion of Earth around the Sun. Owing to the Sun
gravitational potential only, the energy of a neutron varies from $9.12$ eV
to $9.43$ eV between the aphelion and the perihelion. This corresponds to an
absolute shift of $\eta $ of about $1.7$ meV per day. Of course, the full
time-dependence of $\eta$ could also have other origins. For instance, the
relative neutron motion with respect to the unknown matter distribution in
the hidden brane. Nevertheless, it seems unlikely that our own solar system
is ''close'' enough to a similar mass distribution (in the other brane) to
induce a time-dependence on a timescale of the order of one day or one year.
In this context, the most likely time-dependence will be induced by the
Earth motion around the Sun, such that $\Delta \eta \approx 0.31$ eV on one
year. From Eq. (\ref{Proba}), one can then expect a relative variation of
the measured probability $p$ about $\Delta p/p\sim 2\Delta \eta /\eta $. If
the neutron oscillation between branes is detected and presents an annual
dependence through $\Delta p$, since we can estimate $\Delta \eta$ we can
therefore expect specifying the value of $\eta$.

\section{Measurements and analysis}

\label{sec3}

\subsection{Limit of the swapping probability}

\label{analysis}

\begin{table}[t]
\begin{center}
\begin{tabular}{lllll}
Experiment & $\tau_{\mathrm{st}} [\mathrm{s}]$ & $\gamma [\mathrm{Hz}]$ & $%
p_{\mathrm{max}} \times 10^6$ &  \\ \hline\hline
Mampe \textit{et al} \cite{Mampe} & 713 & 17 & $16 \pm 2$ &  \\
Nesvizhevsky \textit{et al} \cite{Nesvizhevsky} & 875 & 4 & $5 \pm 1$ &  \\
Arzumanov \textit{et al} \cite{Arzumanov} & 780 & 9 & $17 \pm 4$ &  \\
Serebrov \textit{et al} \cite{Serebrov} & 873 & 2.6 & $6 \pm 1$ &  \\
Pichlmaier \textit{et al} \cite{Pichlmaier} & 771 & 13 & $13 \pm 1$ &
\end{tabular}
\end{center}
\caption{Summary of UCN storage experiments, with measured UCN storage time
and wall collision rate taken from the original literature. The maximum loss
probability at wall collision is derived for each experiment. }
\label{StorageExp}
\end{table}

In a typical experiment, ultracold neutrons are stored in a bottle, with a
mean wall collision rate $\gamma $ which is typically in the range from 1~Hz
to 100~Hz. The number of stored neutrons decays by following a nearly
exponential law with a decay time $\tau _{\mathrm{st}}$. This storage time $%
\tau _{\mathrm{st}}$ is measured by counting the remaining neutrons after a
storage period of variable duration. The inverse of the storage time is the
sum of the neutron beta decay rate and the loss rate due to wall collisions:
\begin{equation}
\frac 1{\tau _{\mathrm{st}}}=\Gamma _\beta +\Gamma _{\mathrm{loss}}+\gamma p
\label{taust}
\end{equation}
Here we have separated the contribution from the normal loss rate $\Gamma _{%
\mathrm{loss}}$ (due to inelastic scattering of neutrons at the surface for
example) and that corresponding to a disappearance in the other brane $%
\gamma p$.

The purpose of trap experiments is to measure the beta decay lifetime of the
neutron $\tau_n = 1/\Gamma_{\beta}$, by extrapolating the storage time to
the ideal case where there is no extra losses. The extrapolation procedure
is far from trivial and, as stated in the Particle Data Group compilation
\cite{PDG}, the different experimental results are contradictory. Here we
reinterpret some of the performed experiments to provide an upper limit on
the exotic disappearance probability $p$.

Since the extrapolation procedure is in question, we will not try to account
for the normal losses. Instead we shall attribute all the losses to the
exotic phenomenon, and treat the obtained value for the loss as an upper
limit on $p$: $\Gamma_{\mathrm{loss}} + \gamma p < \gamma p_{\mathrm{lim}}$.
This way the presented analysis is not concerned by the present dispute
about the neutron decay lifetime (the normal losses certainly satisfy $%
\Gamma_{\mathrm{loss}} > 0$!).

The idea of the analysis is to compare the neutron decay rate measured in
the absence of brane swapping and the storage time of stored ultracold
neutrons where swapping occur at a rate $\gamma p$. When measuring $\tau _n$
with the beam method, one really measures the beta decay channel: one
measures the absolute proton activity of a cold neutron beamline. Byrne
\textit{et al} \cite{Byrne} measured the rate of proton production of a well-defined section of a cold neutron beam at the ILL and have reported a
neutron lifetime of $\tau _n=889.2\pm 4.8\,\mathrm{s}$. Using the same
improved technique, Nico \textit{et al} \cite{Nico} measured $\tau
_n=886.3\pm 3.4\,\mathrm{s}$. The two independent results can be combined:
\begin{equation}
\tau _n=887.3\pm 2.8\,\mathrm{s}  \label{taun}
\end{equation}

\begin{figure}[t]
\centerline{\ \includegraphics[height=10.5cm]{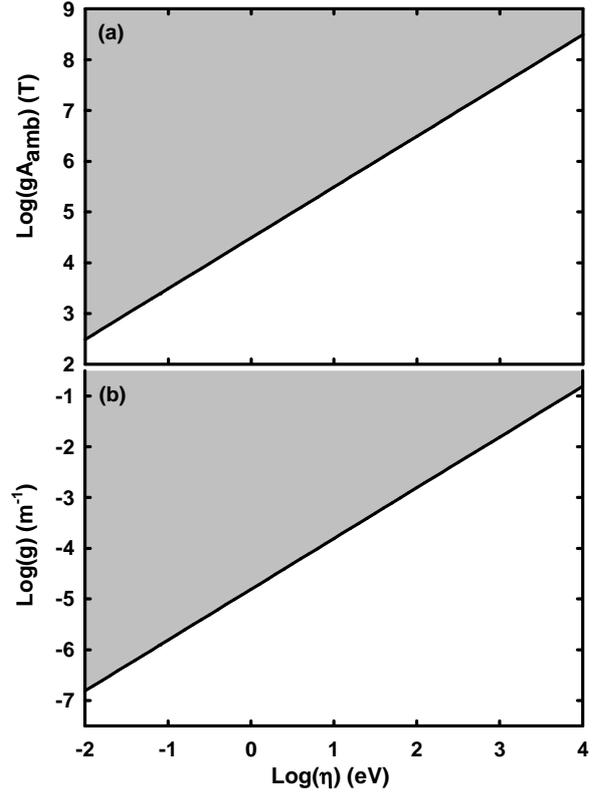}}
\caption{Experimental limits against the confinement energy $\eta \hbar$ of $%
gA_{amb}$ (a) and $g$ (b) for $A_{amb}=2\times 10^9$ T$\cdot $m. Grey
domains are excluded.}
\label{fig1}
\end{figure}

Next we consider the UCN trap experiments \cite
{Mampe,Arzumanov,Serebrov,Pichlmaier} performed at the ultracold neutron
beamline PF2 at the Institut Laue Langevin, using UCN traps coated with
fluorinated polyether oil (Fomblin). We also consider the experiment \cite
{Nesvizhevsky} performed at the Saint Petersburg Institute of Nuclear
Physics using a trap coated with solid Oxygen. In these experiments, the
geometry of the UCN trap could be changed (thus changing $\gamma$) and
several measurements of $\tau_{\mathrm{st}}$ are done corresponding to
different $\gamma$ values. Table \ref{StorageExp} shows the results
extracted from the publications using only the data with the best storage
time. In the last column, the maximal allowed probability for a neutron to
escape in the other brane $p_{\mathrm{lim}}$ is extracted for each storage
experiment using (\ref{taust}) and the pure beta decay lifetime value (\ref
{taun}).

This analysis allows us finally to conclude
\begin{equation}
p < 7 \times 10^{-6}\quad \quad \mathrm{(at\,95\%\,C.L.)}  \label{result}
\end{equation}

This conservative bound could be made even more robust by considering also $%
\tau _n$ value extracted from magnetic trapping of UCNs when available.
Indeed, when neutrons are magnetically trapped they cannot swap to the other
brane, thus the lifetime value measured with magnetic traps could eventually
be combined with the beam average (\ref{taun}).

\subsection{Constraints on $g$ and $\eta$}

\label{constraint}

As a consequence of Eqs. (\ref{Proba}) and (\ref{result}), it becomes possible
to constrain the values of the coupling constant $g$ between the two
braneworlds, and also the environmental potential $\eta $. Fig. 1a shows $%
gA_{amb}$ limits as a function of the confining energy $\eta $. Similarly,
Fig. 1b shows the limit for $g$ and varying $\eta $ (assuming $%
A_{amb}=2\times 10^9$ T$\cdot $m). The resulting constraint is much better
than our previous assessment in earlier works where the upper limit of $g$
was obtained from considerations about millicharged particles \cite{PLB}. In
this case, it was shown that the upper constraint on millicharge $q=\pm
\varepsilon e$ is given by $\varepsilon =$ $(g/2m_e)^2$ ($e$ is the absolute
value of the electron charge and $m_e$ the electron mass). From Ref. \cite
{PLB}, we get $g<3\times 10^{10}$ m$^{-1}$ from the millicharge constraint $%
\varepsilon <4.1\times 10^{-5}$ \cite{milliold}. Even with more recent and
stringent constraints \cite{MirrorPosi2,milli}, we get $g<9\times 10^7$ m$^{-1}$ from $%
\varepsilon <3\times 10^{-10}$ (see Berezhiani and Lepidi in Ref. \cite
{milli}). As a result, ultracold neutron experiments appear as a relevant
approach to constrain $g$ (Some other possibilities are not considered for obvious reasons\footnote{Other ways could be expected to constrain $g$ but are
not relevant for now. Constraint from the primordial nucleosynthesis \cite
{Nucleo} cannot be obtained since the value of a primordial cosmological
magnetic vector potential cannot be assessed at present. In addition,
constraint from disappearance of bound neutrons in nuclei \cite{boundneu}
would need for a rigorous expression of the swapping probability in this
case. There is no simple relation between the disappearance probabilities of
free neutron and neutron in nuclei and this complex topic is far beyond the
scope of the present paper \cite{boundprob}.} \cite{Nucleo,boundneu}).

\section{Further experiments}

\label{further}

\begin{figure}[t]
\centerline{\ \includegraphics[height=6cm]{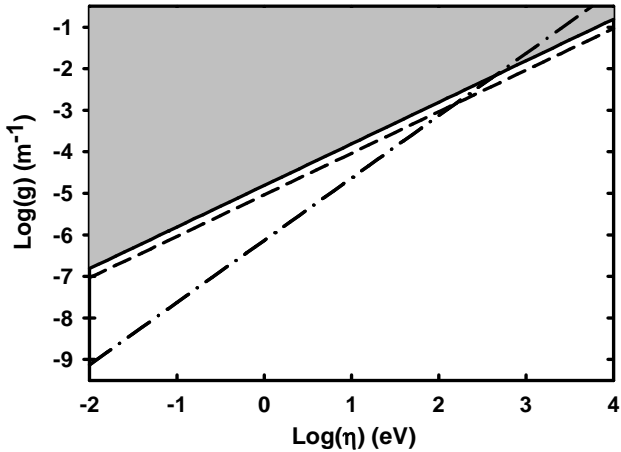}}
\caption{Solid line: Experimental limits for the coupling constant $g$
against the confinement energy $\eta \hbar$. Dashed-doted line: expected limits
for a yearly experiment. Dashed line: expected limits for a resonant
experiment (laser-induced). Grey domain is excluded.}
\label{fig2}
\end{figure}

\subsection{Variable-collision-rate experiment}

\label{Variable}

The most simple further experiment relies on a device with variable
geometry, such that the ratio \textit{volume/surface} of the vessel can be
controlled. The collision rate $\gamma $ then varies as the geometrical
ratio. Obviously, the broader the sampling of $\gamma $ is, the more
accurate the knowledge of $p$. Moreover, for a given collision rate, the
statistical sensitivity on the loss probability is
\begin{equation}
\Delta p=\frac 1\gamma \frac{\Delta \tau }{\tau _n^2}  \label{sensi}
\end{equation}
It is then relevant to increase the value of $\gamma $ to decrease $\Delta p$%
. For instance, the statistical sensitivity of present UCN storage
experiments is about $\Delta \tau \approx 1$~s when measuring the neutron
lifetime $\tau _n\approx 900$~s. With a current wall collision rate of $%
\gamma =10$~Hz, this translates into $\Delta p\approx 10^{-7}$. By contrast,
with $\gamma =100$ Hz, $\Delta p\approx 10^{-8}$ is a fair reachable
sensitivity. Such an experimental approach can be accepted with confidence.
Indeed, the lowest considered frequencies of the swapping probability are
about $10^{13}$ Hz in relation to $\eta $ (see section \ref{model}). This is
far to the highest expected collision rates, and then Eqs. (\ref{6}) and (%
\ref{Proba}) are still valid.

\subsection{Yearly experiment}

\label{yearly}

As suggested in section \ref{envpot}, it would be highly beneficial to
perform the experiment on a long timescale (typically one year) to take into
account the motion of Earth around the Sun and the related time-dependence
of $\eta$. Such a motion should lead to a time-modulation of the swapping
probability $p$. If one can detect such a modulation it would be a strong
indication that matter swapping really occurs. Indeed, with such an
experiment it could be possible to discriminate the exotic losses due to
matter swapping from normal losses, which are not time-dependent, in the
neutron trap. In addition, if the neutron swapping between braneworlds is
detected and presents an annual dependence through $\Delta p$, since one can
estimate $\Delta \eta $ one could then assess the value of $\eta$.

Moreover, a clear benefit of any long time experiment would be to constrain
the unknown value of $\Delta p$ against time. This allows to improve the
constraints on $gA_{amb}$ (or $g$) against $\eta $ by contrast with
experiments related to the upper limit of $p$ only. In the present paper, in
order to underline the relevance of the yearly experiment, we can just
suggest a test value for the limit of $\Delta p$. As shown from (\ref{sensi}%
), $\Delta p\approx 10^{-7}$ when $\gamma =10$~Hz. Since the uncertainty
varies as $1/\sqrt{N}$, $\Delta p\approx 10^{-8}$ is a fair reachable upper
experimental limit\footnote{%
At present, values about $50$ neutrons per cm$^3$ are reachable. $500$
neutrons per cm$^3$ are expected soon. In addition, a one year experiment
would allow to obtain at least ten times more of measurements. As a
consequence, the number of events $N$ would be hundred times greater than
now.} for a one year experiment.

If we assume that $\Delta \eta \approx 0.31$ eV on one year (see section \ref
{envpot}) from the above values and since $\Delta p/p\sim 2\Delta \eta /\eta$%
, we can then further improve the constraint on $gA_{amb}$ (or $g$) against $%
\eta$. Fig. 2 shows the expected limits for the coupling constant $g $
against $\eta$ for a yearly experiment (dashed-doted line). The result is
compared with that found previously in section \ref{analysis}. It becomes
obvious that a yearly experiment allows getting a much better assessment of $%
g$ when $\eta \leq 2(p/\Delta p)\Delta \eta$. The rule is clearly that the
weaker the relative uncertainty of $p$ is, the better the constraint is.
From the above benchmark values, one deduces that the yearly experiment
could offer a better estimation when $\eta \leq 434$ eV by contrast with the
present experimental constraints (solid line). For the lower values of $\eta$
considered here, the gain could reach two orders of magnitude.

\subsection{Laser-induced matter swapping}

\label{induced}

In previous works \cite{Reso, fcomb}, it has been suggested that a rotative
magnetic vector potential could be considered instead of a static one. In
that case a resonant swapping occurs with Rabi oscillations given by:
\begin{equation}
P=\frac{4\Omega ^2}{(\eta -\omega )^2+4\Omega ^2}\sin ^2\left( (1/2)\sqrt{%
(\eta -\omega )^2+4\Omega ^2}t\right)  \label{RabiR}
\end{equation}
where $\omega $ is the angular frequency of the magnetic potential. Equation
(\ref{RabiR}) shows that a resonant matter exchange between branes occurs
whenever the magnetic vector potential rotates with an angular frequency $%
\omega =\eta $. Of course, in this case, we do not consider an astrophysical
field, and the rotative magnetic potential is supplied by an electromagnetic
wave.

In a recent paper \cite{fcomb}, it has been suggested that a set of coherent
electromagnetic pulses (a frequency comb) could artificially induce the
swapping of a neutron into a hidden braneworld. The neutron swapping rate $%
\Gamma$ is then given by \cite{fcomb}:
\begin{equation}
\Gamma =K\frac{f_r\tau ^2NI}{\eta ^2}g^2  \label{18}
\end{equation}
where $f_r$ is the frequency of repetition of the pulses, $\tau $ is the
pulse duration, $N$ is the number of pulses felt by neutrons, $I$ is the
intensity of the pulse. In the above expression, $\eta $ is given in eV, $%
f_r $ in GHz, $I$ in PW$\cdot$cm$^{-2}$, $\tau $ in fs, and $K=\mu
^2/(50c\varepsilon _0e^2)\approx 2.74\times 10^{-14}$ (in the relevant units).

From the uncertainty on (\ref{taun}), a relevant criterion to confirm the
reality of this effect is to achieve $\Gamma \geq \Delta \tau _n/\tau
_n^2=3\times 10^{-3}\cdot \Gamma _\beta $. In Fig. 2, one shows the expected
limits for such a resonant experiment assuming that the laser frequency can
be continuously tuned (dashed line). This figure was derived assuming the
following values for the frequency comb source: $\tau =1$ ps, $f_r=100$ GHz,
$N=150$, $I=10^8$ PW$\cdot$cm$^{-2}$. These values are quite usual for some
frequency comb sources (see references in \cite{fcomb}). Though the
intensity is more specific to certain pulsed sources (see discussions in
Refs. \cite{Laserint}), such a value could be achieved for a frequency comb
as well. It is striking that the theoretical limit (dashed line) is very
close to the present experimental limit (solid line). Due to the regular
improvement of laser sources, we expect that such a resonant experiment will
become relevant in the next decades by contrast to passive experiments to
explore the space of parameters $(\eta ,g)$.

\section{Conclusion}

Using results from performed experiments, we have assessed an upper limit on
the probability for a stored ultracold neutron to disappear into another
braneworld. This limit has been used to constrain the parameters of the
brane model introduced in recent theoretical works \cite{BraneNC,PLB,Reso}, which had shown the possibility of matter
exchange between two braneworlds invisible to each other. We have discussed
the sensibility of further experiments to probe the existence of a
neighboring brane through an annual study. It is also suggested that a
laser-induced matter swapping towards a hidden braneworld could be tested in
the next decades.

\section*{Note added in proof}
During the process of publication of this paper, we learned about the work of Berezhiani and Nesti \cite{BerNesti} whose results could be reminiscent of ours. Nevertheless, the results of Ref. \cite{BerNesti} are independent of ours. Indeed, our present work rests on a different physical approach, which is fully specific to the braneworld concept \cite{BraneNC}. In addition, our present model allows for enhanced induced matter exchange between parallel braneworlds by artificial means \cite{Reso,fcomb}. In this case, the efficiency of the matter swapping rate is not limited and is proportional to the intensity of the available laser sources (see Eq. (\ref{18})) \cite{fcomb}.

\section*{Acknowledgement}

The authors are grateful to Gia Dvali and Vladimir Tretyak for their useful
advices and comments.


\begin{thebibliography}{99}
\bibitem{Review}  K. Akama, Lect. Notes Phys. 176 (1983) 267, arXiv:hep-th/0001113; \\ V.A. Rubakov, M.E. Shaposhnikov, Phys. Lett. 125B (1983) 136;\\ M. Pavsic, Phys. Lett. 116A (1986) 1, arXiv:gr-qc/0101075; \\ P. Horava, E. Witten, Nucl. Phys. B460 (1996) 506, arXiv:hep-th/9510209;\\ A. Lukas, B.A. Ovrut, K.S. Stelle, D. Waldram, Phys. Rev. D 59 (1999) 086001, arXiv:hep-th/9803235; \\ L. Randall, R. Sundrum, Phys. Rev. Lett. 83 (1999) 3370, arXiv:hep-ph/9905221; \\ R. Davies, D.P. George, R.R. Volkas, Phys. Rev. D 77 (2008) 124038, arXiv:0705.1584 [hep-ph]; \\ Y.-X. Liu, L.-D.
Zhang, L.-J. Zhang, Y.-S. Duan, Phys. Rev. D 78 (2008) 065025,
arXiv:0804.4553 [hep-th].

\bibitem{Pheno}  N. Arkani-Hamed, S. Dimopoulos, G. Dvali, Phys. Lett. B 429 (1998) 263, arXiv:hep-ph/9803315;\\ I. Antoniadis, N. Arkani-Hamed, S.
Dimopoulos, G. Dvali, Phys. Lett. B 436 (1998) 257,
arXiv:hep-ph/9804398;\\ N. Arkani-Hamed, S. Dimopoulos, G. Dvali, Phys. Rev. D 59 (1999) 086004,
arXiv:hep-ph/9807344.

\bibitem{bbb}  G. Dvali, G. Gabadadze, Phys. Lett. B 460 (1999) 47, arXiv:hep-ph/9904221.

\bibitem{HEP}  D. Hooper, S. Profumo, Phys. Rep. 453 (2007) 29, arXiv:hep-ph/0701197.

\bibitem{Gravi}  J. Chiaverini, S.J. Smullin, A.A. Geraci, D.M. Weld, A.
Kapitulnik, Phys. Rev. Lett. 90 (2003) 151101,
arXiv:hep-ph/0209325; \\ Y. Shtanov, A. Viznyuk, Class. Quant. Grav. 22 (2005) 987,
arXiv:hep-th/0312261.

\bibitem{Mirrornu}  R. Foot, R.R. Volkas, Phys. Rev. D 52 (1995) 6595, arXiv:hep-ph/9505359; \\ Z. Berezhiani, R.N.
Mohapatra, Phys. Rev. D 52 (1995) 6607, arXiv:hep-ph/9505385.

\bibitem{Tunnel}  S.L. Dubovsky, V.A. Rubakov, P.G. Tinyakov, Phys. Rev. D 62 (2000) 105011,
arXiv:hep-th/0006046;\\ S.L. Dubovsky, JHEP 0201 (2002) 012, arXiv:hep-th/0103205;\\ C. Ringeval, P. Peter, J.-P. Uzan, Phys. Rev. D 65
(2002) 044016, arXiv:hep-th/0109194.

\bibitem{Mirrorphot1}  R. Foot, A.Yu. Ignatiev, R.R. Volkas, Phys. Lett. B 503 (2001) 355,
arXiv:astro-ph/0011156.

\bibitem{Mirrorphot2}  S. Abel, B. Schofield, Nucl. Phys. B685 (2004) 150, arXiv:hep-th/0311051.

\bibitem{Mirrorphot3}  S.A. Abel, J. Jaeckel, V.V. Khoze, A. Ringwald, Phys. Lett. B 666 (2008) 66,
arXiv:hep-ph/0608248.

\bibitem{Mirrorneut}  Z. Berezhiani, Eur. Phys. J. C 64 (2009) 421,
arXiv:0804.2088 [hep-ph]; \\ Z. Berezhiani, L. Bento, Phys. Rev. Lett. 96
(2006) 081801, arXiv:hep-ph/0507031.

\bibitem{MirrorPosi} P. Crivelli, A. Belov, U. Gendotti, S. Gninenko, A. Rubbia, JINST 5 (2010) P08001, arXiv:1005.4802 [hep-ex];\\
R. Foot, S.N. Gninenko, Phys. Lett. B 480 (2000) 171, arXiv:hep-ph/0003278;\\ S.N. Gninenko, N.V. Krasnikov, A.
Rubbia, Phys. Rev. D 67 (2003) 075012, arXiv:hep-ph/0302205.

\bibitem{MirrorPosi2} A. Badertscher, et al., Phys. Rev. D 75 (2007) 032004, arXiv:hep-ex/0609059.

\bibitem{hidden}  G. Dvali, M. Redi, Phys. Rev. D 80 (2009) 055001,
arXiv:0905.1709 [hep-ph].

\bibitem{BraneNC}  M. Sarrazin, F. Petit, Phys. Rev. D 81 (2010) 035014,
arXiv:0903.2498 [hep-th].

\bibitem{PLB}  F. Petit, M. Sarrazin, Phys. Lett. B 612 (2005) 105, arXiv:hep-th/0409084.

\bibitem{Reso}  M. Sarrazin, F. Petit, Int. J. Mod. Phys. A 22 (2007) 2629, arXiv:hep-th/0603194.

\bibitem{fcomb}  M. Sarrazin, F. Petit, Phys. Rev. D 83 (2011) 035009, arXiv:0809.2060
[hep-ph].

\bibitem{IJMPA}  M. Sarrazin, F. Petit, Int. J. Mod. Phys. A 21 (2006) 6303,
arXiv:hep-th/0505014.

\bibitem{neule} I. Antoniadis, et al., Comp. Rend. Phys. 12 (2011) 755;\\
    V.V. Nesvizhevsky, G. Pignol, K.V. Protasov, Phys. Rev. D 77 (2008) 034020, arXiv:0711.2298 [hep-ph].

\bibitem{neuexp1}  I. Altarev, et al., Phys. Rev. D 80 (2009) 032003, arXiv:0905.4208 [nucl-ex];\\ A.P. Serebrov, et al.,
Nucl. Instrum. Meth. A 611 (2009) 137, arXiv:0809.4902 [nucl-ex];%
\\ A.P. Serebrov, et al., Phys. Lett. B 663 (2008) 181, arXiv:0706.3600 [nucl-ex];\\ G. Ban, et al., Phys. Rev. Lett. 99 (2007) 161603, arXiv:0705.2336 [nucl-ex].

\bibitem{neuexp2}  S. Baessler, V.V. Nesvizhevsky, K.V. Protasov, A.Yu.
Voronin, Phys. Rev. D 75 (2007) 075006, arXiv:hep-ph/0610339.

\bibitem{DGS}  G. Dvali, G. Gabadadze, M. Shifman,
Phys. Lett. B 497 (2001) 271, arXiv:hep-th/0010071;\\
G. Dvali, G. Gabadadze, M. Porrati,
Phys. Lett. B 485 (2000) 208, arXiv:hep-th/0005016;
\\S.L. Dubovsky, V.A. Rubakov, Int. J.
Mod. Phys. A 16 (2001) 4331, arXiv:hep-th/0105243.

\bibitem{bigrav}  Z. Berezhiani, F. Nesti, L. Pilo, N. Rossi, JHEP 0907 (2009) 083, arXiv:0902.0144 [hep-th].

\bibitem{Ignatovich} M. Utsuro, V. K. Ignatovich, \textit{Handbook of Neutron Optics}, Wiley-VCH (2010).

\bibitem{vectpot}  R. Lakes, Phys. Rev. Lett. 80 (1998) 1826;\\ J. Luo, C.-G. Shao, Z.-Z. Liu, Z.-K. Hu, Phys. Lett. A 270 (2000) 288.

\bibitem{vectpot2}  A.S. Goldhaber, M.M. Nieto, Phys. Rev. Lett. 91 (2003) 149101, arXiv:hep-ph/0305241.

\bibitem{GalMag}  E. Asseoa, H. Sol, Phys. Rep. 148 (1987) 307.

\bibitem{GalMag2}  A. Neronov, I. Vovk,
Science 328 (2010) 73, arXiv:1006.3504 [astro-ph.HE].

\bibitem{PDG}  Particle Data Group, K. Nakamura, et al., J. Phys. G 37 (2010) 075021.

\bibitem{Byrne}  J. Byrne, et al., Europhys. Lett. 33 (1996) 187.

\bibitem{Nico}  J.S. Nico, et al., Phys. Rev. C 71 (2005) 055502, arXiv:nucl-ex/0411041.

\bibitem{Mampe}  W. Mampe, P. Ageron, C. Bates, J.M. Pendlebury, A.
Steyerl, Phys. Rev. Lett. 63 (1989) 593.

\bibitem{Nesvizhevsky}  V.V. Nesvizhevsky, et al., Sov. Phys. JETP 75 (1992) 405;\\
A.G. Kharitonov, et al., Nucl. Inst. Meth. A 284 (1989) 98.

\bibitem{Arzumanov}  S. Arzumanov, et al., Phys. Lett. B 483 (2000) 15.

\bibitem{Serebrov}  A.P. Serebrov, et al., Phys. Rev. C 78 (2008) 035505, arXiv:nucl-ex/0702009.

\bibitem{Pichlmaier}  A. Pichlmaier, V. Varlamov, K. Schreckenbach, P.
Geltenbortd, Phys. Lett. B 693 (2010) 221.

\bibitem{milliold}  A.A. Prinz, et al., Phys. Rev. Lett. 81 (1998) 1175, arXiv:hep-ex/9804008.

\bibitem{milli}  Z. Berezhiani, A. Lepidi, Phys. Lett. B 681 (2009) 276,
arXiv:0810.1317 [hep-ph]; \\ A. Melchiorri, A. Polosa, A. Strumia, Phys. Lett. B 650
(2007) 416, arXiv:hep-ph/0703144;\\ S. Davidson, S. Hannestad, G. Raffelt, JHEP 0005 (2000) 003, arXiv:hep-ph/0001179.

\bibitem{Nucleo}  G.J. Mathews, T. Kajino, T. Shima, Phys. Rev. D 71 (2005) 021302, arXiv:astro-ph/0408523;\\ J.S. Nico, W.M. Snow, Ann. Rev. Nucl. Part. Sci. 55 (2005) 27, arXiv:nucl-ex/0612022;\\ Z. Berezhiani, L. Bento,
Phys. Lett. B 635 (2006) 253, arXiv:hep-ph/0602227.

\bibitem{boundneu}  SNO Collaboration, S.N. Ahmed, et al., Phys. Rev. Lett. 92 (2004) 102004,
arXiv:hep-ex/0310030; \\ KamLAND collaboration, T. Araki, et al., Phys. Rev.
Lett. 96 (2006) 101802, arXiv:hep-ex/0512059.

\bibitem{boundprob}  V.I. Nazaruk, Int. J. Mod. Phys. E 20 (2011) 1203, arXiv:1004.3192
[hep-ph];\\ V. Kopeliovich, I. Potashnikova, Eur. Phys. J. C 69 (2010) 591, arXiv:1005.1441 [hep-ph];%
\\ V.I. Nazaruk, Phys. Rev. C 58 (1998) 1884, arXiv:hep-ph/9810354.

\bibitem{Laserint}  A. Ipp, J. Evers, C.H. Keitel, K.Z. Hatsagortsyan, Phys.
Lett. B 702 (2011) 383, arXiv:1008.0355 [physics.ins-det];\\T.
Heinzl, J. Phys.: Conf.
Ser. 198 (2009) 012005;\\S.-W. Bahk, et al., Opt. Lett. 29 (2004) 2837.

\bibitem{BerNesti}Z. Berezhiani, F. Nesti, arXiv:1203.1035 [hep-ph].

\end{thebibliography}
\end{document}